\documentclass[12pt]{article}
\usepackage{amsmath}
\usepackage{float}
\usepackage{url}
\usepackage{subcaption}
\expandafter\def\csname ver@subfig.sty\endcsname{}
\usepackage{comment}
\usepackage{algorithm2e}
\usepackage{amssymb}

\usepackage{comment}
\usepackage{graphicx} 
\usepackage{array}
\usepackage[english,french]{babel}
\usepackage[utf8]{inputenc}
\usepackage{float} 
\usepackage{enumitem}
\usepackage{multicol}
\usepackage{shuffle}
\usepackage{bbold}
\usepackage[tikz]{bclogo}
\usepackage{xcolor}
\usepackage{stmaryrd}
\usepackage{amsthm}
\usepackage{dsfont}
\usepackage{tkz-graph}
\usepackage{tikz}
\usepackage{tikz-3dplot}
\usepackage{hyperref}
\usepackage{appendix}
\usepackage[T1]{fontenc}

\usepackage{graphicx}
\usepackage{color}

\usepackage[left=2.5cm,right=2.5cm,top=2.5cm,bottom=2.5cm]{geometry}
\newtheorem{theorem}{Theorem}[section]
\newtheorem{definition}{Definition}[section]

\begin{document}


\begin{center}
{\Large
	{\sc  Time topological analysis of EEG using signature theory}
}
\bigskip

Stéphane Chrétien $^{1}$ \& Ben Gao $^{2}$ \& \underline{Rémi Vaucher} $^{3}$ \& Astrid Thébault Guiochon $^{4}$
\bigskip

{\it
$^{1}$ Laboratoire ERIC, Université Lyon 2, France, stephane.chretien@univ-lyon2.fr\\
$^{2}$ Halias Technologie, Grenoble, France, ben.gao@halias.fr\\
$^{3}$ Halias Technologies \& Laboratoire ERIC, université Lyon 2, remi.vaucher@halias.fr\\
$^4$ Laboratoire EMC, Université Lyon 2, a.thebaultguiochon@univ-lyon2.fr}
\end{center}
\bigskip


{\bf R\'esum\'e.} La détection d'anomalies dans les signaux multivariés est une tâche de première importance dans de nombreuses disciplines (épidémiologie, finance, sciences cognitives et neuro-sciences, cancérologie etc). Dans cette optique, l'analyse topologique des données (TDA) offre une batterie d'invariants "de forme" qui peuvent être exploités pour la mise en ouvre d'un schéma de détection efficace. Notre contribution consiste à étendre les construction présentées dans \cite{chretienleveraging} sur la construction de complexes simpliciaux à partir des Signatures des signaux et leur capacités prédictives, plutôt que de l'utilisation d'une distance générique comme dans \cite{petri2014homological}. La théorie des Signatures est une nouvelle thématique en Machine Learning \cite{chevyrev2016primer} issue des travaux récents sur les notions de Chemins Rugueux (Rough Paths) développés par Terry Lyons et son équipe \cite{lyons2002system} sur la base du formalisme introduit par Chen \cite{chen1957integration}. Nous explorons en particulier la détection des changements de topologie, sur la base du suivi de l'évolution de la persistance homologique et des nombres de Betti associés au complexe introduit dans \cite{chretienleveraging}. Nous appliquons nos outils pour l'analyse de signaux cérébraux de type EEG afin de détecter des phénomènes précurseurs aux crises d'épilepsies. 

{\bf Mots-cl\'es.} Analyse de données topologique, Séries Temporelles, Statistiques Appliquées

\medskip

{\bf Abstract.} 
Anomaly detection in multivariate signals is a task of paramount importance in many disciplines (epidemiology, finance, cognitive sciences and neurosciences, oncology, etc.). In this perspective, Topological Data Analysis (TDA) offers a battery of "shape" invariants that can be exploited for the implementation of an effective detection scheme. Our contribution consists of extending the constructions presented in \cite{chretienleveraging} on the construction of simplicial complexes from the Signatures of signals and their predictive capacities, rather than the use of a generic distance as in \cite{petri2014homological}. Signature theory is a new theme in Machine Learning \cite{chevyrev2016primer} stemming from recent work on the notions of Rough Paths developed by Terry Lyons and his team \cite{lyons2002system} based on the formalism introduced by Chen \cite{chen1957integration}. We explore in particular the detection of changes in topology, based on tracking the evolution of homological persistence and the Betti numbers associated with the complex introduced in \cite{chretienleveraging}. We apply our tools for the analysis of brain signals such as EEG to detect precursor phenomena to epileptic seizures.

{\bf Keywords.} Topological and Geometric Data Analysis, Time Series, Applied Statistics

\bigskip\bigskip


\section{Introduction}

Topological data analysis \cite{chazal2021introduction} is a rapidly growing field that has emerged recently, based on the intriguing observation that data come with shape-like properties. In general, existing topological structures that are built from data, such as Cech or Vietoris Rips complexes, make essential use of a metric that may not be fully suitable because inherited from the space where our data are imbedded instead of the geodesic distance on the manifold where our data truly live. In a recent paper \cite{chretienleveraging}, we proposed a novel way of buiding simplicial complexes based on performance in prediction rather than distance. The main tool for performing prediction is the signature transform \cite{chevyrev2016primer}, that easily extract meaningful features from multidimensional signals. In this setting, each node represents a component of the signal and a node belongs to a simplex if the signature of all the nodes from the simplex accurately explains the considered node, statistically speaking. Similarly, a face belongs to a simplex if the signature of the signals represented by the face is accurately predicted by the adjacent faces of various orders in the simplex. In this approach, Signatures faithfully represent faces and simplices together with a natural orientation and make the topological construction better motivated and more statistically meaningful.   From a computational viewpoint, selecting faces that explain another one using their respective signatures can be done efficiently using the LASSO, in the spirit of the celebrated method proposed in \cite{meinshausen2006high} for estimating graphical models. More recently, \cite{petri2014homological} solved this problem more precisely by considering a signal correlation matrix. This method allows to create high-dimensionnal structures.

In the present paper, we examine a more refined aspect of our topological construct: the dynamic evolution of the topology as a function of time as complexes undergo potential structural transformations at specific change points in time, reflecting the apperance of certain phenomena. In the area of neuroscience, this approach will be instrumental for detecting change points at which electroencephalograms reflects known "neuroscientific" behaviors.

\subsection{Signature of rough paths (in a nutshell)}

Consider a $d$-dimensionnal path $X=(X^1,...,X^d):\mathbb{R}\rightarrow \mathbb{R}^2$. In the following, we will note $S_I^{(k)}(X)$ the $k$-th degree signature applied to $X$ on an time interval $I=[a,b]$. The \textbf{signature} of $X$ is given by the tensors sequence $S_I^{(k)}(X)\in\mathbb{R}^{\underbrace{d\times d\times \dots \times d}_{k\text{ times}}}$ for all $k\in\mathbb{N}$. We will use the notion of truncated signature of order $K$ to define the sequence of signature tensor $S_I^{(k)},k\leq K$. These tensors are given, for $\{i_1,...,i_k\}\subset\{1,...,k\}$ by:
\begin{align*}
    \left( S_I^{(k)}(X)\right)_{i_1,...,i_k}=S_I^{i_1,...,i_k}(X)=\idotsint\limits_{a<t_1<t_2<\dots<t_k<b}dX_{t_1}^{i_1}dX_{t_2}^{i_2}\dots dX_{t_k}^{i_k}
\end{align*}

The signature of a rough path serves as a powerful geometric feature extractor. To compute the signature of a multidimensional signal that is discretely sampled, we must consider linear interpolation between each consecutive time point, thus invoking Chen's theorem:

\begin{theorem}[Chen's identity]
    Consider $X:[a,b]\rightarrow\mathbb{R}^d$ and $Y:[b,c]\rightarrow\mathbb{R}^d$. Define:
    \begin{align*}
        X\ast Y & = \left\{ \begin{array}{ccl}
             X_t,& \text{ si }t\in[a,b]  \\
             Y_t-Y_0+X_b,& \text{ si }t\in[b,c]
        \end{array}\right.
    \end{align*}
    as the \textbf{concatenation} of $X$ and $Y$. Then:
    \begin{align*}
        S_I^{(k)}(X\ast Y) & = S_I^{(k)}(X)\otimes S_I^{(k)}(Y)
    \end{align*}
\end{theorem}

To center/rescale the signals, one has to consider the following properties
\begin{itemize}
    \item For any constant $\gamma\in\mathbb{R}^d$, $S_I^{(k)}(X+\gamma)=S_I^{(k)}(X)$
    \item For any constant $\lambda\in\mathbb{R}$, $S_i^{(k)}(\lambda X)=\lambda^kS_I^{(k)}(X)$
\end{itemize}

Furthermore, to ensure a "pseudo" unicity (because it is computationally infeasible to create the whole signature) we will consider the next result:

\begin{theorem}
    If $\exists i\in\{1,...,k\}$ such that $X^i$ strictly monotonic, then $S_I(X)$ uniquely defines $X$.
\end{theorem}

Finally, defining $\mathcal C_{\text{mon}}^{1-var}(I,\mathbb R^d)$ the space of  $d$-dimensionnal signals (linearly interpolated on a subdivision $D(I)$ of $I$) with  $|X|_{1-var}=\sup\limits_{t_i\in D(I)}\sum\limits_i d(X_{t_i},X_{t_{i+1}})<\infty$ and such that $\exists i\in\{1,\dots,d\}$ with $X^i$ strictly monotonic, then the signature transform $S_I:\mathcal C_{\text{mon}}^{1-var}(I,\mathbb{R}^d)\rightarrow S_I(\mathcal C_{\text{mon}}^{1-var}(I,\mathbb{R}^d))$ is an homeomorphism. \cite{friz2010multidimensional}

\subsection{Simplicial complex}

Consider a set of $d$ vertices $V=\{X^1,...,X^d\}$.

\begin{definition}
    \begin{itemize}
    \item For $k<d$, a $k$-simplex $\sigma_k$ is a $n+1$ set and all its subsets \cite{schenck2022algebraic}.
    \item A geometric realization of $\sigma_k$ is given by the convex hull $E_k$ of its $n+1$ points (eventually embedded) in $\mathbb{R}^d$ such that $\dim(E_k)=k$ (so $d\geq k$)
    \end{itemize}
\end{definition}

With this tool, one can build a new structure on $V$\cite{chazal2021introduction}.

\begin{definition}
    An abstract simplicial complex $\mathcal C$ on a finite vertex set $V$ is a collection $\{\sigma_k$, $k<d\}$ of simplices such that:
    \begin{itemize}
        \item $v\in\mathcal{C}$ if $v\in V$
        \item $\tau\subseteq \sigma\in\mathcal C\Rightarrow \tau \in \mathcal C$  
    \end{itemize}
    The \textbf{dimension} of $\mathcal{C}$ is the highest $k$ such there exists a $k$-simplex in $\mathcal{C}$.
\end{definition}
A simplex $\sigma \in \mathcal C$ is called a \textbf{face} of $\mathcal C$.
The upcoming definition is the most important for our algorithm:
\begin{definition}
    Consider $k \in N^*$ and $\sigma$ a $k$-simplex in a simplicial complex $\mathcal C$. Its \textbf{link} $\text{Lk}(\sigma,\mathcal C )$ in $\mathcal C$ is the set of all faces $\tau\subset\mathcal C$ such that:
    \begin{itemize}
        \item $\sigma\cap \tau = \varnothing$
        \item $\sigma\cup\tau$ is a face of $\mathcal{C}$
    \end{itemize}
\end{definition}
\textbf{Remarks:}
\begin{itemize}
    \item The link of a vertex is called \textbf{neighborhood} in graph theory.
    \item In the following algorithm, for a vertex $v$ and a fixed $k$, we build the subset of all the $k$-simplex in $\text{Lk}(v,\mathcal{C})$ that we call $k$-dimensional Link (for simplicity). 
\end{itemize}

\subsection{Some tools for simplicial complex analysis.}
The main goal of this section is to introduce some useful topological invariants, the first of these being the \textbf{Betti numbers}\cite{chazal2021introduction}. Informally, each $b_k$ count the number of $k+1$-dimensional holes in $\mathcal{C}$. In our experiments, we focus on $2$-dimensional complexes, and thus only consider $b_0$ (the number of connected components) and $b_1$ (the number of 2-dimensional holes).\\

The second invariant that we used is the \textbf{Persistence Diagram}. To get a numerical summary of this, we compute the notion of \textbf{Persistence Entropy}\cite{rucco2016characterisation,chintakunta2015entropy}. A more robust notion of Persistence Summary presented in \cite{chung2022persistence} could be put to work, but the implementation raises much more complex issues that we will not address in the present work.

\begin{definition}
    Consider a $k$-dimensional simplicial complex $\mathcal C$. A \textbf{filtration} over $\mathcal C$ is a sequence $(\mathcal C_i)_{0\leq i\leq n}$ of simplicial complexes such that:
    \begin{itemize}
        \item For all $1\leq i\leq j\leq n$, $\mathcal C\subset \mathcal C_j$
        \item $\mathcal C_n =\mathcal C$
    \end{itemize}
\end{definition}

A natural way of building a filtration is by defining an increasing (non necessary strictly) sequence of time of arrival $(b_i)_{1\leq i\leq n}$ for each simplex in $\mathcal C$. In Cech complex, the time of arrival correspond to the radius $r$ at wich the simplex appear in $\mathcal C$.

\begin{definition}
   Consider a simplicial complex $\mathcal{C}$ and his persistence diagram $DP$
    \begin{align*}
        DP &= \{[b_i,d_i), i\in\{1,...,\#{\mathcal{C}}\}\}
    \end{align*}
    with $b_i$ and $d_i$ respectively birth and death times for sub-simplex of $\mathcal{C}$.\\
    Define 
    \begin{align*}
        p_i=\frac{b_i-d_i}{\sum\limits_{j}{b_j-d_j}}
    \end{align*}
    \textbf{Persistence entropy} (PE) is given by the Shannon entropy of $\{pi\}$:
    \begin{align*}
        PE = -\sum\limits_{i}p_i\log(p_i)
    \end{align*}
\end{definition}

Simplistically , the persistence entropy measures the similarity (or dissimilarity) between closure speed of $k$-dimensional holes: high level PE indicates that all holes are filled at same speed, in opposition to a PE close to 0. Since a quickly filled hole is likely just noise, PE quantifies how many significatives sub-structures lies in $\mathcal C$.\\

Beginning with section 2, we will consider these structures and invariants as evolving as a function of time: for any timestamp $t\in I$, we will consider the simplicial complex $\mathcal{C}(t)$, its associated betti numbers $b_0(t),...,b_n(t)$ and its related persistence entropy denoted by $PE(t)$.\\

\textbf{Remark:} many other features may be extracted from the filtration we introduce for $\mathcal{C}$. More will be studied in a long version of this contribution.

\subsection{Our simplicial complex construction algorithm}

\subsubsection{The algorithm}
Our algorithm build the simplicial complex on $X = \{X^1,...,X^d\}$ by constructing the $k$-dimensional link (iteratively on $k$) of a channel $X^i$. Here is a simplified version of the algorithm \cite{chretienleveraging}.\\

\begin{algorithm}
\caption{Sequential construction of a simplicial complex on $(X^1,...,X^d)$}\label{alg:two}
\KwData{Fix the degree $deg$ of signature and a computation interval $I$.}
\KwResult{The simplicial complex of interaction among each time series.}
\For{$k$ from 1 to $d-1$}{
  \For{$i$ from 1 to $d$}{
   Compute $S^{deg}(\tilde{X^i})$ (where $\tilde{X^i}(t)=(t,X^i(t))$).\;
  For every word $i_1...i_k$ of $\{1,...,d\}\setminus\{i\}$, compute $S_I^{deg}(\tilde{X}^{i_1...i_k})$ where $\tilde{X}^{i_1...i_k}(t)=(t,X^{i_1},...,X^{i_k})$\;
  Predict $S_I^{deg}(\tilde{X^i})$ from $\left\{S_I^{deg}(\tilde{X}_{i_1...i_{k}})\right\}_{{i_1...i_k}\in(\{1,...,d\}\setminus\{i\})^k}$ with LASSO\;
  \eIf{$R^2>0,67$}{Select all non-zero $\beta_{i_1...i_k}$}
  {Fix $\beta_{i_1...i_k}=0$ for all $i_1...i_k$.
    }
  \For{Every $\beta_{i_1\dots i_k}\neq 0$}{ Create the simplex whose vertices are $\{X^i,X^{i_1},...,X^{i_k}\}$}
  }
  }
\end{algorithm}

\subsubsection{The induced filtration:}
In order to retrieve the persistence diagram, we need to construct a filtration and so birth time.

\begin{itemize}
    \item Take all the weights $(\beta_\sigma)_{\sigma\in\mathcal C}$ attached to each simplex.
    \item Create $b_\sigma = 1-\frac{\beta_\sigma}{\sum\limits_{\sigma\in\mathcal C}}\beta_\sigma$.
    \item For each $\tau,\sigma\in\mathcal C$ such that $\tau\subset \sigma$ and $b_\tau>b_\sigma$, then fix $b_\tau = b_\sigma$.
\end{itemize}

This (non necessarily strictly) increasing sequence ensures that a highly significative simplex appears early in the filtration. In the following section, we show how to exploit this filtration.

\section{Empirical study}

We performed a set of computational experiements using  EEG signals from the CHB-MIT Scalp Database \cite{goldberger2000physiobank}. The main benefit of using real-world signals is to test the stability of our algorithm against a uncontrolled noise. 

\subsection{Importance of hyperparameter values}
We will evaluate the impact of the choice of the main parameters based on a one-hour trajectory, averaged every second. The signals reflect the occurence of an epilepsy seizure which t lasts less than one minute (on the interval $[49',51']$). 
We focus our exploration on the LASSO parameters and the sliding Window size for the computation of the Signature. For the sake of simplicity, we will specify the maximal dimension of $\mathcal{C}(t)$ as equal to 2.

\begin{description}

    \item[LASSO parameters] For every channel $X^i$, we solve the following minimization problem:
    \begin{align}    \min\limits_{\beta\in\mathbb{R}}\|S_I^{deg}(\tilde{X}^i)-\sum\limits_{j\neq i}\beta_jS_I^{deg}(\tilde{X}^j)\|_2^2+\lambda_1\sum\limits_{j\neq i}|\beta_j|
    \end{align}

    LASSO allows us to select which signatures (and then by homeomorphism which $X^j$) lies in the neighbourhood of $X^i$. This selection depends entirely on $\lambda_1$. However, this minimization only creates the 1-dimensional neighbourhood of $X^i$. to creates the 2-dimensional NH, we solve:
    \begin{align}              \min\limits_{\beta\in\mathbb{R}}\|S_I^{deg}(\tilde{X}^i)-\sum\limits_{j_1,j_2\neq i,j_1\neq j_2 i}\beta_{j_1j_2}S_I^{deg}(\tilde{X}^{j_1j_2})\|_2^2+\lambda_2\sum\limits_{j\neq i}|\beta_{j_1j_2}|
    \end{align}
    This naturally brings a second parameter.

    Experimental results not included in the present manuscript but available in the Appendix show that fixing $\lambda_1$ and varying $\lambda_2$ does not change $b_1$ nor the $PE$ trajectory. Then, fixing $\lambda_2$ and varying $\lambda_1$ does not give the same result: on the one hand, similarly to the precedent case, persistence entropy is not much impacted by how $\lambda_1$ varies. On the other hand, $b_1$ 's trajectory becomes noisier when one enforces sparsity in the LASSO.

    \item[Sliding window size for computing the Signature :] At each time $t$ we build a simplicial complex $\mathcal{C}_t$, and for a path $P$ defined on $I$, we compute $S_{[t-L,t]}^{k}(P)$ (under the condition $[t-L,t]\subset I$). Length $L$ of $I$ is then a hyperparameter that can impact the topological structure of $\mathcal C(t)$. First results show that choosing $L$ is key to safe detection of the sought for patterns.

\end{description}

\subsection{Experiments with multiple EEG trajectories.}

In the proposed experiments, we focus on a multivariate EEG signal sampled during 15 consecutives hours, and that includes 3 seizures. The method's hyperparameters are specified as follow:
\begin{itemize}
    \item $\lambda_1=1$ and $\lambda_2=1$
    \item $L=50$.
\end{itemize}
The main goal is to retrieve pre-critical and/or critical behaviour based on the Betti numbers and the Persistence Entropy, our main topological invariants of interests. According to \cite{navarro2010seizure},

\begin{description}
    
\item[Pre-critical behaviour:] are caracterized by "area partitioning" (loss of connectivity between neuronal areas, loss of synchrony). Our main hope is that this behaviour is reflected in the trajectory of $b_1$: a loss of connectivity should result in poorly interconnected cortical areas, resulting in several stabilized 2-dimensional holes. An expected dynamics is then to observe the steady growth of $b_1$ in time. It would be relevant to extend the dimension to $k=3$ in order to see if this fact is confirmed with $b_2(t)$. Another idea would be to caracterise every hole in $\mathcal C$ and track the stable ones among them.

\item[Critical behaviour:] Neuronal populations will abruptly synchronise throughout the duration of the crisis, leading to hypersynchrony before returning to a more 'local' level of synchrony.
\end{description}

In Figure \ref{b1_multicurves} below, one observes the global behavior of $b_1$ and $PE$ on 1 hour of time (second trajectory). We computed an average measure on a clean trajectory (the green line) for both $b_1$ and $PE$. The grey area represent the interval $[x(t)-\hat{\sigma_x},x(t)+\hat{\sigma_x}$ with $x=b_1$ or $PE$ and $\hat{\sigma_x}$ computed on $[t-h,t]$.

\begin{figure}[H]
    \centering
        \includegraphics[width=.8\linewidth]{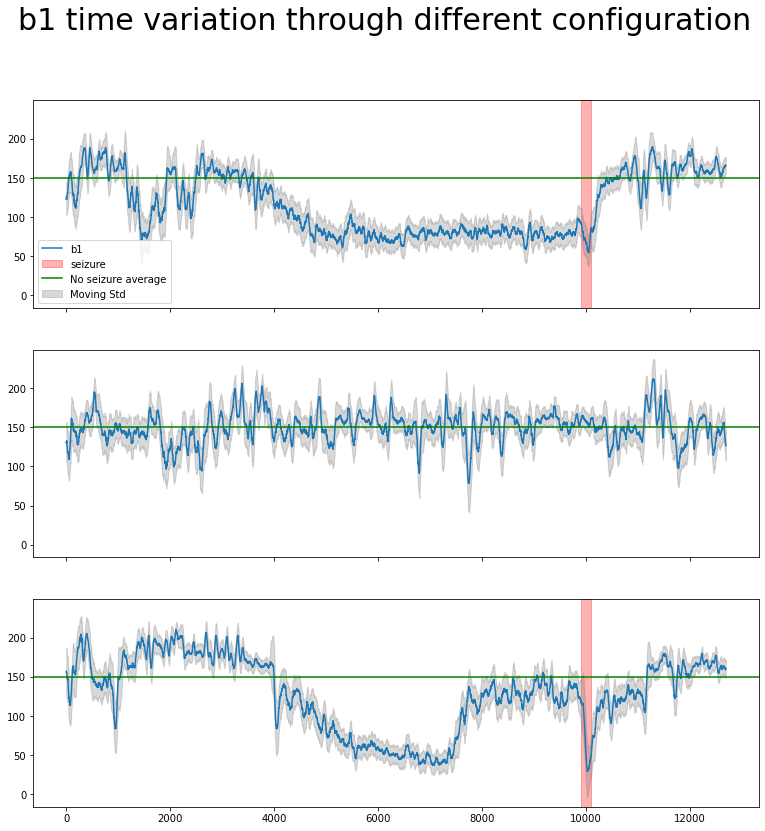}
        \caption{$b_1(t)$ on 3 trajectories}
        \label{b1_multicurves}
\end{figure}
\begin{figure}[H]
    \centering
        \includegraphics[width=.8\linewidth]{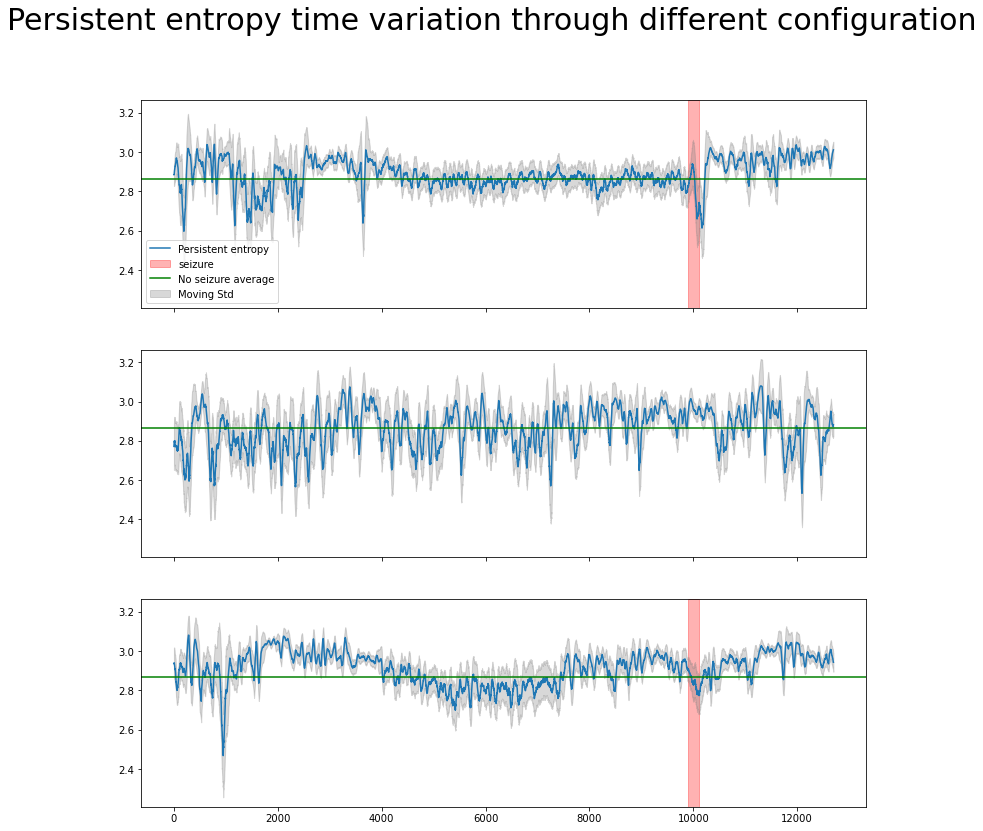}
        \caption{$PE(t)$ on 3 trajectories}
        \label{pe_multicurves}
\end{figure}

\begin{description}
    \item[General behavior (second trajectory):] We see a volatility around the average value along time, with small perturbations that we cannot explain. The volatility goes from 100 to 200 for $b_1$ and from 2.6 to 3 for $PE$.
    \item[Critical behavior (red area in first and third trajectory):] The critical behavior is caracterized by an abrupt diminution to 50 in $b_1$'s trajectory  (more connected structure in the simplicial complex before getting back to near the average value. Unfortunately, it cannot really be distinguished from $PE$.
    \item[Pre-critical behavior (6000 times before critical):] The precritical behavior is characterised by a steep decrease of $b_1$ (between 50 and 100) on the same time interval for the two seizure trajectories. An sudden increase in noticeable just before the next seizure, whilst remaining under the average value.\\
    Heuristically, we can observe a decrease of the $PE$'s standard deviation during the precritical stage. We plan to extend this study in a future work. 
\end{description}

\subsection{Discussion}

Many improvements can still be made to the proposed methodology, in particular (i) using Knockoff filters to guarantee that the faces are selected with confidence, (ii) using SLOPE instead of the LASSO, (iii) using higher-order sub-complexes, etc. One may also use the path $(b_1(t),PE(t))$ and its signature as well for extracting more sensitive detection pipelines. 

\bibliographystyle{plain}
\bibliography{biblio}

\appendix

\section{Appendix}
\subsection{Impact of the hyperparameters for $b_1 (t)$}
\subsubsection{$\lambda_i$:}
\begin{figure}[h]
    \centering
    \begin{minipage}[t]{0.4\textwidth}
        \centering
        \includegraphics[width=\textwidth]{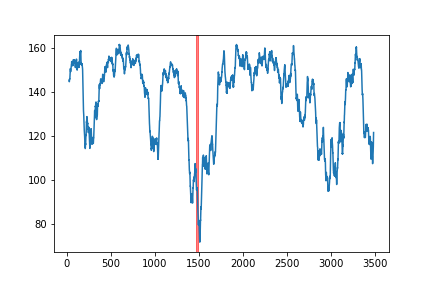}
        \caption*{a) $\lambda_1=\lambda_2=1$.}
    \end{minipage}
    \hfill
    \begin{minipage}[t]{0.4\textwidth}
        \centering
        \includegraphics[width=\textwidth]{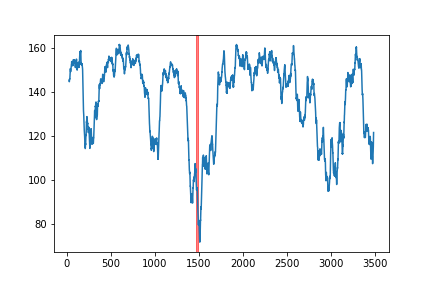}
        \caption*{$\lambda_1=1$, $\lambda_2=10^4$}
    \end{minipage}
    \hfill
    \begin{minipage}[t]{0.4\textwidth}
        \centering
        \includegraphics[width=\textwidth]{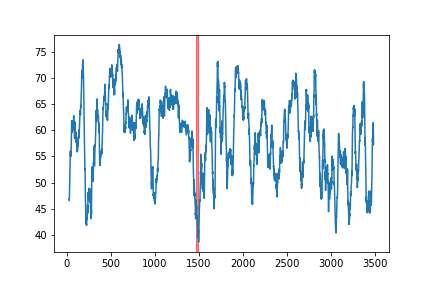}
        \caption*{c) $\lambda_1=10^4$, $\lambda_2=1$}
    \end{minipage}
    \hfill
    \begin{minipage}[t]{0.4\textwidth}
        \centering
        \includegraphics[width=\textwidth]{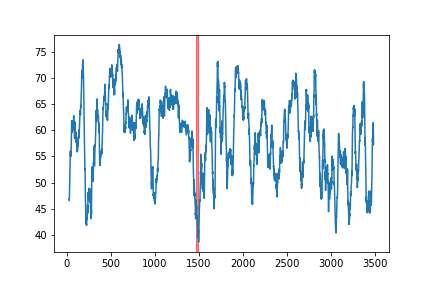}
        \caption*{c) $\lambda_1=10^4$, $\lambda_2=10^4$}
    \end{minipage}
    \caption{Study of the effect of $\lambda_1$ and $\lambda_2$ on $b_1(t)$ (with a seizure in red)}.
    \label{b1lamb}
\end{figure}
As can be observed, when  $\lambda_2$ increases, $b_1(t)$ remains the same on the full interval. On the other hand, a noisier trajectory results from changing the value of $\lambda_1$. One suspects that  only the 1-dimensional edges that are not subsets of 2-dimensional faces (dis)appear as we move $\lambda_i$.
\newpage
\subsubsection{$L$:}
\begin{figure}[h]
    \centering
    \begin{minipage}[t]{0.4\textwidth}
        \centering
        \includegraphics[width=\textwidth]{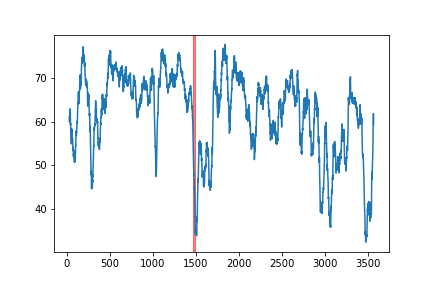}
        \caption*{a) $L=20s$.}
    \end{minipage}
    \hfill
    \begin{minipage}[t]{0.4\textwidth}
        \centering
        \includegraphics[width=\textwidth]{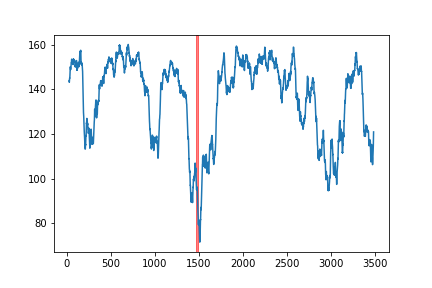}
        \caption*{$L=100s$}
    \end{minipage}
    \hfill
    \begin{minipage}[t]{0.4\textwidth}
        \centering
        \includegraphics[width=\textwidth]{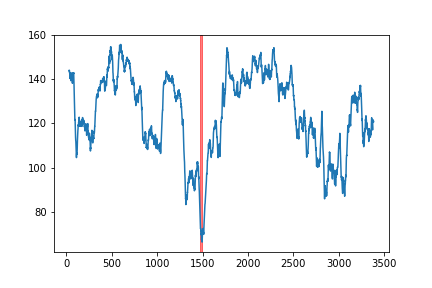}
        \caption*{c) $L=200s$}
    \end{minipage}
    
    \caption{Study of the influence of $L$ on $b_1(t)$ (with a seizure in red)}.
    \label{b1L}
\end{figure}
As can be expected from varying $L$, keeping $L$ at a low level make the signal more noisy. Surprisingly, the amplitude changes as $L$ is increased, implying that either more edges are created or less 2-dimensional faces are created (or both). This fact will be studied more precisely in a longer version of the paper.

\newpage
\subsection{Parameters study for $PE(t)$}
\subsubsection{$\lambda_i$:}

\begin{figure}[h]
    \centering
    \begin{minipage}[t]{0.4\textwidth}
        \centering
        \includegraphics[width=\textwidth]{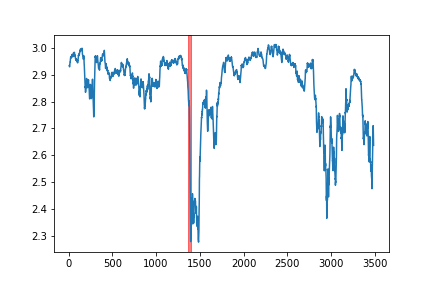}
        \caption*{a) $\lambda_1=\lambda_2=1$.}
        \label{fig:img_a}
    \end{minipage}
    \hfill
    \begin{minipage}[t]{0.4\textwidth}
        \centering
        \includegraphics[width=\textwidth]{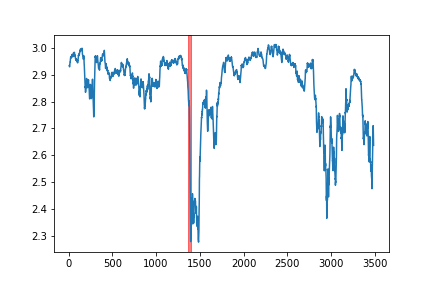}
        \caption*{$\lambda_1=1$, $\lambda_2=10^4$}
    \end{minipage}
    \hfill
    \begin{minipage}[t]{0.4\textwidth}
        \centering
        \includegraphics[width=\textwidth]{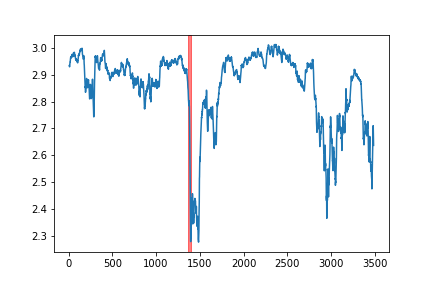}
        \caption*{c) $\lambda_1=10^4$, $\lambda_2=1$}
    \end{minipage}
    \hfill
    \begin{minipage}[t]{0.4\textwidth}
        \centering
        \includegraphics[width=\textwidth]{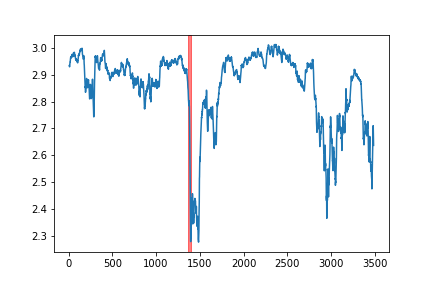}
        \caption*{c) $\lambda_1=10^4$, $\lambda_2=10^4$}
    \end{minipage}
    \caption{Study of the effect of $\lambda_1$ and $\lambda_2$ on $PE(t)$ (with a seizure in red)}.
    \label{PElamb}
\end{figure}

It is clear from the numerical experiments that persistence entropy is not impacted by varying  $\lambda_1$. Related to the variations of $b_1$, one can deduce that the created edges did not influence persistence entropy.

\newpage

\subsubsection{$L$:}
\begin{figure}[h]
    \centering
    \begin{minipage}[t]{0.4\textwidth}
        \centering
        \includegraphics[width=\textwidth]{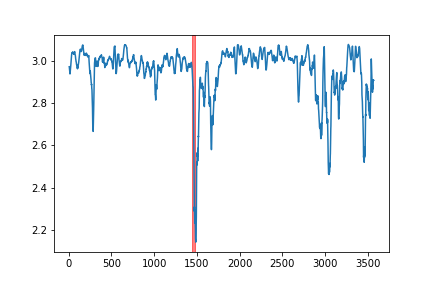}
        \caption*{a) $L=20s$.}
    \end{minipage}
    \hfill
    \begin{minipage}[t]{0.4\textwidth}
        \centering
        \includegraphics[width=\textwidth]{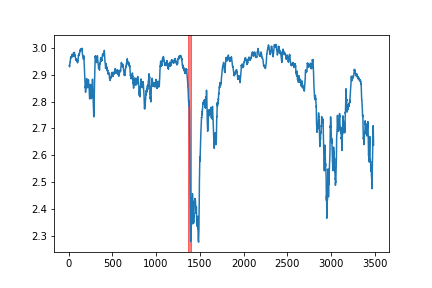}
        \caption*{$L=100s$}
    \end{minipage}
    \hfill
    \begin{minipage}[t]{0.4\textwidth}
        \centering
        \includegraphics[width=\textwidth]{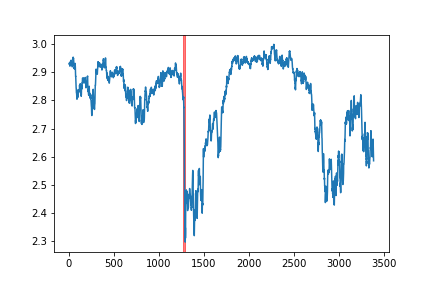}
        \caption*{c) $L=200s$}
    \end{minipage}
    
    \caption{Study of the effect of $L$ on $PE(t)$ (with a seizure in red)}.
    \label{PEL}

\end{figure}
Clearly, lower level of the window size ensures the stability of persistent entropy. With this stability comes shorter events (like at $t=300$ or $t=3000$). Higher levels of window size brings longer events, and earlier distinguishable (as the small event at $t=500$ show) 

\end{document}